\documentclass[twocolumn,showpacs,preprintnumbers,amsmath,amssymb]{revtex4}
\usepackage[dvips]{graphicx}
\usepackage{color}

\def\ket#1{  \left\vert  #1   \right\rangle   }

\begin{document}

\title{Gossip Algorithms in Quantum Networks}

\author{Michael Siomau}

\email{siomau@nld.ds.mpg.de}

\affiliation{Physics Department, Jazan University, P.O.~Box 114,
45142 Jazan, Kingdom of Saudi Arabia and \\
Network Dynamics, Max Planck Institute for Dynamics and
Self-Organization (MPIDS), 37077 G\"{o}ttingen, Germany}

\date{\today}

\begin{abstract}
Gossip algorithms is a common term to describe protocols for
unreliable information dissemination in natural networks, which are
not optimally designed for efficient communication between network
entities. We consider application of gossip algorithms to quantum
networks and show that any quantum network can be updated to optimal
configuration with local operations and classical communication.
This allows to seed-up -- in the best case exponentially -- the
quantum information dissemination. Irrespective of the initial
configuration of the quantum network, the update requiters at most
polynomial number of local operations and classical communication.
\end{abstract}


\maketitle

\section{Introduction}
\label{intro}

Real-world networks are complex: natural social and brain networks
as well as artificial technological and computer networks exhibit
non-trivial structural features, which make complete simulation of
the network dynamics practically impossible \cite{Dorog:03}. Complex
non-stationary structure of modern artificial networks becomes a
serious obstacle in the design of optimal protocols for information
dissemination in such networks. Inspired by a natural way of rumor
spreading in social networks, gossip algorithms \cite{Shah:08} give
a simple strategy for distributed and robust information
dissemination in a network of unknown structure. These algorithms
have found prominent applications in sensor, peer-to-peer and social
networks.

Quantum networks \cite{Kimble:08} will be the next generation of
complex structures for communication and advanced information
processing \cite{Dowling:03}. Due to quantum superposition and
nonlocality \cite{Nielsen:00}, quantum networks exhibit a number of
structural and dynamical features that classical networks lack,
among those are teleportation \cite{Bennett:93}, quantum walks
\cite{Venegas:12} and entanglement percolation
\cite{Acin:07,Siomau:16a} to name just a few. Recently we showed
that with local operations and classical communication (LOCC)
\cite{Nielsen:00} one may change connectivity of a given quantum
network and simulate complex entanglement graphs on a simple
underlying quantum network \cite{Siomau:16b}. The structural
modifications may radically improve the network capacity for
information dissemination and performance of corresponding
protocols, such as gossip algorithms.

In this paper we consider the problem of optimal information
dissemination in quantum networks and analyze performance of gossip
algorithms on the networks. As intuition suggests, the network where
any pair of vertices is connected with an edge offers the most
favorable conditions for information dissemination. Such a network
is represented with a complete graph. We show that any quantum
network represented with a connected graph, i.e. where any two
vertices can be connected with a path of edges, may be updated to
the complete graph using just polynomial number of LOCC. The update
allows to dissimilate information by means of quantum teleportation
\cite{Nielsen:00}, thus radically improving the performance of the
gossip algorithms on quantum networks.

This work is structured as follows. In the next section, we briefly
describe classical gossip algorithms for single- and multi--piece
information dissemination and introduce the quantities of interest,
such as conductance, $k$-conductance and $\varepsilon$-dissemination
time. For a more detailed and mathematically rigorous treatment we
suggest an excellent review by Shah \cite{Shah:08}. In
Section~\ref{sec:3}, we show how to improve the performance of
gossip algorithms on quantum networks by LOCC. For sparse quantum
networks the improvement in the information dissemination time due
to the update is exponential, but still requires only polynomial
number of LOCC. We conclude in Section~\ref{sec:4}.

\section{Classical Gossip Algorithms}
\label{sec:2}

From the structural viewpoint a network is a graph $G = (V,E)$
defined by sets of its vertices $V$ and edges $E$. The set
$V=\{1,...,n\}$ consists of a finite countable number of $n$
vertices. The edges represent connections between the vertices $E
\subset V \times V$. The graph is called undirected if for any
$(i,j) \subset E$, $(j,i) \subset E$ is also true. Here we impose no
constrains on the direction of information dissimilation, hence
consider only undirected graphs.

Information dissimilation on a graph may be studied with a discrete
random walk technique, which requires definition of a $n \times n$
non-negative valued probability transition matrix $P = [P_{ij}]$,
where $P_{ij}$ is the probability of information dissemination from
vertex $i$ to $j$. Through the transition matrix, we may define an
auxiliary function named conductance $\Phi(P)$, which characterizes
the information dissemination capacity of a graph of particular
configuration of vertices and edges. For symmetric $P$ -- which is
the case for undirected graphs --  the conductance is defined as
\cite{Shah:08}
\begin{equation}
\label{conductance}
 \Phi(P) \, = \, \min_{S \subset V:|S| \leq n/2}
\; \frac{\sum_{i\subset S, \, j\subset S^c} P_{ij}}{|S|} \, ,
\end{equation}
where $S$ is the set of nodes that possess the information, while
$S^c$ is the set of those that doesn't. The conductance is
completely defined by the transition matrix of a graph, thus tells
us how easy the information can be conducted through the graph.
Also, the conductance is independent on a particular information
dissemination protocol to be implemented on the graph.

A related to the conductance auxiliary function is $k$-conductance,
which minimizes (\ref{conductance}) for $k\leq n/2$, i.e.
\begin{equation}
\label{k-conductance}
 \Phi_k(P) \, = \, \min_{S \subset V:|S| \leq k}
\; \frac{\sum_{i\subset S, \, j\subset S^c} P_{ij}}{|S|} \, .
\end{equation}
Using the $k$-conductance, we may also define the mean conductance
$\hat{\Phi}(P)$ as
\begin{equation}
\label{mean-conductance}
 \hat{\Phi}(P) \, = \, \sum_{k=1}^{n-1} \frac{k}{\Phi_k(P)} \, .
\end{equation}

In the following we will focus on two particular graphs: the
complete graph, where each pair of nodes is connected with an edge,
and the ring graph, where nodes are placed on a circle with edges
between nearest neighbors only. These two graphs are chosen for
comparison because of their radical difference in the capacity for
information dissemination. With the probability matrix $P_{ij} =
1/n$ for all $i$ and $j$, the complete graph has the best possible
capacity to disseminate information, i.e. $\Phi(P) = O(1)$ and
$\hat{\Phi}(P)= O(n^2 \log n)$, where $O(..)$ is the standard
notation for asymptotic upper bound. The ring graph with the
probability matrix $P_{ii} = 1/2$ and $P_{ij} = 1/4$ for $i\neq j$,
in contrast, has the strongest constrain for information
dissemination leading to $\Phi(P) = O(1/n)$ and $\hat{\Phi}(P)=
O(n^3)$.

Analyzing gossip algorithms we will be interested in the value
called $\varepsilon$-dissemination time $T(\varepsilon)$. This value
gives us time by which all nodes have the information with
probability at least $1-\varepsilon$. The definition of the
$\varepsilon$-dissemination time depends on the algorithm, thus will
be given in the next sections for single- and multi-piece
dissemination strategies separately. Our goal is to estimate the
$\varepsilon$-dissemination time through the conductance, allowing
general treatment of the algorithm efficiency for any graph
structure.

\subsection{Single-Piece Dissemination}
\label{subsec:2.1}

Let an arbitrary vertex $\upsilon \in V$ has a piece of information
that it wishes to spread to all the other vertices as quickly as
possible. Let $S(t) \subset V$ denotes the set of vertices that have
the information at time $t$, which is also assumed to be discrete.
At each time step, each vertex $i$ contacts {\it at most one} of its
neighbors $j$ with probability $P_{ij}$. If either $i$ or $j$ has
the information at $t-1$, then both vertices have it at time $t$.

For the single-piece dissemination algorithm, the
$\varepsilon$-dissemination time is defined as
\begin{equation}
T_1 (\varepsilon) = \sup_{\upsilon \in V} \inf \{t: {\rm Pr} \left(
S(t) \neq V | S(0) = {\upsilon} \right) \leq \varepsilon \}.
\end{equation}
The right hand side of this definition accounts for the maximal time
at which the set $S(t)$ is inequivalent to $V$ with probability no
greater then $\varepsilon$, assuming that initially the set $S(t=0)$
consisted of a single vertex $\upsilon$.

The $\varepsilon$-dissemination time for the single-piece
dissemination algorithm may be expressed through the conductance
(\ref{conductance}) as \cite{Shah:08}
\begin{equation}
 \label{single-piece time}
T_1 (\varepsilon) = O \left( \frac{\log n + \log
\varepsilon^{-1}}{\Phi(P)} \right).
\end{equation}
This expression tells us explicitly how the
$\varepsilon$-dissemination time depends on the structure of
underlying network, i.e. on its conductance. For the complete graph
the $\varepsilon$-dissemination time is given by $T_1^{c}
(\varepsilon) = O(\log n)$, which is the upper bound for
single-piece dissemination algorithm performance in any network. For
the ring graph the $\varepsilon$-dissemination time is exponentially
larger comparing to the previous case, i.e. $T_1^{r} (\varepsilon) =
O(n \log n)$. It is important to note that information dissemination
on a ring can be performed as fast as $O(n)$ by setting a simple
intuitive rule, for example, 'always send information to the left
neighbor'. But, gossip algorithms has no account for network
structure, which is the key for their universality. Moreover, the
gossip algorithms on a ring are just logarithmically slower then the
intuitive strategy, which is practical.

\subsection{Multi-Piece Dissemination}
\label{subsec:2.2}

In contrast to single-piece dissemination algorithm, where just a
single vertex has the information initially, in multi-piece
dissemination each vertex wants to spread its own information to all
the other vertices as quickly as possible. Let $M = \{ m_1,...,m_n
\}$ denotes the set of messages at time $t=0$. As before each vertex
contacts at most one of its neighbors at each time step. During the
contact, the vertices exchange all information they don't have. The
$\varepsilon$-dissemination time is defined as
\begin{eqnarray}
T_M (\varepsilon) = & & \nonumber
\\[0.1cm]
& & \hspace*{-1cm} \inf \{t: {\rm Pr} \left( \bigcup_{i=1}^n S_i(t)
\neq M | S_i(0) = m_i \right) \leq \varepsilon \} \, ,
\end{eqnarray}
i.e. the maximal time at which the information at each vertex is
inequivalent to the initial set $M$ with probability no greater then
$\varepsilon$. The $\varepsilon$-dissemination time is expressed
through the mean conductance (\ref{mean-conductance}) as
\cite{Shah:08}
\begin{equation}
 \label{multi-piece time}
T_M (\varepsilon) = O \left( \frac{\hat{\Phi}(P) \log
\varepsilon^{-1}}{n} \right).
\end{equation}
For the complete graph the $\varepsilon$-dissemination time is given
by $T_M^{c} (\varepsilon) = O(n \log^2 n)$, which is the upper bound
for multi-piece dissemination algorithm performance in any network.
For the ring graph, in contrast, the $\varepsilon$-dissemination
time is exponentially smaller, i.e. $T_M^{r} (\varepsilon) = O(n^2
\log n)$.

\section{Gossip Algorithms in Quantum Networks}
\label{sec:3}

Eqs.~(\ref{single-piece time}) and (\ref{multi-piece time})
unambiguously define performance of gossip algorithms through
conductance (\ref{conductance}) and it's mean
(\ref{mean-conductance}) for any classical network. In the classical
case, there is no option to change the conductance of a network
without addition of physical connections between vertices. In
quantum networks, in contrast, entanglement swapping allows
physically disconnected vertices to become connected with an
entangled state, i.e. an entangled edge, without direct interaction
between the vertices \cite{Nielsen:00}. The entangled state may be
subsequently used for information transmission by means of
teleportation. However, the teleportation of quantum information
also requires classical communication \cite{Bennett:93}. Therefore,
we assume that vertices may freely communicate classically, while
the condition for the gossip algorithm -- namely that each vertex
contacts at most one of its neighbors at each time step -- applies
to the quantum information. This assumption seems reasonable,
because the update of the quantum network with the entangled edges
may be done in advance to gossip algorithm run as we explain below.
In this sense, the update means exploring and improving network
structure for the purpose of future gossip dissemination.

\subsection{Quantum Network Update}
\label{subsec:3.1}

Let us show that any connected graph can be updated to the complete
graph using just polynomial number of LOCC. To do so we need to
estimate the upper bound on LOCC to update a connected graph. An
important local characteristic of any graph is the average degree
$\hat{k} = 2E/V$, i.e. the average number of edges $E$ connected to
vertex $V$. A global characteristic that measures the efficiency of
the information transport in a network is the average path length
\begin{equation}
 \label{SPL}
L_G = \frac{1}{n(n-1)} \sum_{i\neq j} d(v_i, v_j) \, ,
\end{equation}
where $d(v_i, v_j)$ is the shortest distance between vertices $v_i,
v_j \in V$. The graph with the smallest average degree and the
largest average path length is the most constrained for information
dissemination, thus the update of the graph requires maximal number
of LOCC. The ring graph and the 1D chain with $\hat{k} = 2$ and $L_G
= O(1)$ are the desired graphs \cite{Bocc:06} to estimate the upper
bound on LOCC. In the following we will focus on the ring graph
noticing that all considerations remain valid also for the 1D chain.

The procedure of the update begins with creating perfect
entanglement between any pair of physically connected vertices of
the quantum network. To be precise, let us assume that a pair of
qubits in a Bell state \cite{Nielsen:00} is to be distributed
between any pair of physically connected vertices. A perfect Bell
state can be created between two neighboring vertices by exchanging
photons through the edges and, if necessary, purification
\cite{Vidal:99}. The perfect entanglement can be distributed on
arbitrary distance with entanglement swapping \cite{Acin:07}. This
distribution creates a single non-local edge that connects
physically disconnected vertices. Let $\ket{a} = \sum_{i,j} a_{ij}
\ket{ij}$ be a two-qubit state in the computational basis $\{
\ket{0}, \ket{1}\}$. The entanglement of this state can be described
with concurrence \cite{Wootters:98} defined as $C(a)= 2 | \det A |$,
where $A=[a_{ij}]$. After $K$ entanglement swapping operations the
concurrence reads as
\begin{eqnarray}
C_K = & & \sup_M \sum_i 2 | \det \left( A_1 M_1 A_2 M_2 ... A_K M_K
\right)| \, \nonumber
\\[0.1cm]
& & \hspace*{4cm} = \, \prod_{i=1}^K | \det A_i | \, ,
\end{eqnarray}
where $M_i$ for $i=1..K$ are $2\times 2$ matrices that denote the
choice of measurements. The entanglement of the qubit pair after $K$
entanglement swapping remains perfect, i.e. $C_K=1$, iff the initial
entangled states $\ket{a}_i$ where maximally entangled $C(a_i) = 1$.

Summing up all the considerations above, the ring graph can be
updated to the complete graph by sharing multiple copies of perfect
entanglement between neighboring vertices and creating non-local
entanglement edges with (multiple) entanglement swapping. The update
of the ring graph gives the upper bound on LOCC for any connected
graph. In the following we will estimate the bound for the single-
and multi-piece gossip algorithms.

\subsection{Single-Piece Dissemination in Quantum Networks}
\label{subsec:3.2}

Let vertex $\upsilon \in V$ has a piece of quantum information
encoded into a qubit state $\ket{\psi}$ to disseminate among the
others. Because there is just one piece of information to
disseminate, each entangled edge is to be used just once to send the
information. Let us suppose that we have a ring graph with $n$
vertices. For sake of clarity let us assume that $n$ is even,
noticing that the results remain valid for odd $n$. To update the
ring graph to the complete graph we need to create for a chosen
vertex one the longest non-local entanglement edge using $n/2-1$
entanglement swapping operations and doubled number of edges using
$n/2-i$ for $i=2..(n/2-1)$. This procedure is to be repeated for all
vertices excluding duplications. The total number of the non-local
edges to establish is $(n-1)(n/2-1)$. Thus the total number of the
entanglement swapping operations scales as $O(n^3)$
\cite{Siomau:16b}. This is the upper bound on the LOCC for
single-piece dissemination algorithm in any quantum network. As we
showed in Section~\ref{subsec:2.1}, the $\varepsilon$-dissemination
time is exponentially larger in the complete graph $T_1^{c}
(\varepsilon) = O(\log n)$ comparing to the ring graph $T_1^{r}
(\varepsilon) = O(n \log n)$. Thus, the update gives the exponential
benefit in information dissemination time requiring just $O(n^3)$
LOCC.

\begin{figure}
\begin{center}
\includegraphics[scale=0.3]{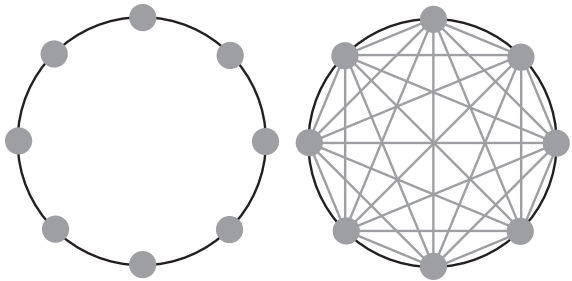}
\caption{A ring graph of eight vertices (left) is updated to the
complete graph (right) exponentially improving the capacity of the
network for information dissemination. Classical edges are shown in
black, while the entangled edges -- in grey.}
 \label{fig}
\end{center}
\end{figure}

Let us consider a ring graph of just eight vertices as shown in
Fig.~\ref{fig}. Starting from an arbitrary vertex, we may
disseminate information through the others at best in seven steps.
If the ring is updated to the complete graph by adding entangled
edges, the same information can be spread in just three steps. This
difference in the information dissemination capacity growth
radically with the network size. For a ring with $n = 2^k$ nodes,
the fastest classical dissemination is possible with $2^k - 1$
steps, while in the updated network, the information can be
disseminated as fast as in $k$ steps. At the same time, the update
requires $n^3$ LOCC, i.e. scales polynomially with the network size.

An interesting aspect of the information dissemination in the
updated network is that this dissemination is secure. In the ring,
each vertex may corrupt information it receives: even though the
information is encoded in quantum states, it is possible to copy the
information partially \cite{Fan:14}. In the updated network, in
contrast, each pair of vertices is connected making the gossip
secure at each step from the other vertices. Overall faster gossip
dissemination reduces the number of potential information
modifications due to previous hosts.

\subsection{Multi-Piece Dissemination in Quantum Networks}
\label{subsec:3.3}

In contrast to the previous case, in multi-piece dissemination
algorithm each vertex has its own qubit state to disseminate, i.e.
$M_{\ket{\psi}} = \{ \ket{\psi}_1,...,\ket{\psi}_n \}$, thus each
edge is to be used $n$ times. But, the non-local entanglement edges
are destroyed after the state teleportation. The simplest way to
overcome this complication is to create the complete graph with $n$
replicas of the non-local edges. This requires just $O(n^4)$ local
operations to update the ring graph, which is still appropriate cost
in our opinion and gives the upper bound on LOCC. The update of the
ring graph allows to improve exponentially the
$\varepsilon$-dissemination time from $T_M^{r} (\varepsilon) = O(n^2
\log n)$ to $T_M^{c} (\varepsilon) = O(n \log^2 n)$.

\section{\label{sec:4} Conclusion}

We suggested a new way to speed-up information distribution in
quantum networks by structural update, which requires at most
$O(n^3)$ and $O(n^4)$ LOCC for single- and multi-piece dissemination
gossip algorithms respectively. Our approach is based solely on
quantum non-locality, i.e. the ability to connect physically
disconnected vertices with entangled states and quantum
teleportation. But, because (classical) gossip algorithms are based
on random walk, we believe that our approach is compatible with
quantum walks \cite{Venegas:12}. Taking into account that gossip
algorithms have applications not only in information dissemination
but in linear and separable function computation \cite{Shah:08}, the
combination of our structural approach with the quantum walks may
lead to new model of quantum computing \cite{Nielsen:00} and quantum
machine learning \cite{Siomau:14} in complex quantum networks.

\begin{acknowledgments}
This work was supported by KACST.
\end{acknowledgments}

\end{document}